\documentstyle[aps,pre,manuscript]{revtex}

\begin{document}

\draft

\title{The noise properties of stochastic processes and entropy production}

\author{Bidhan Chandra Bag, Suman Kumar Banik and 
Deb Shankar Ray{\footnote{e-mail : pcdsr@mahendra.iacs.res.in} }
}

\address{Indian Association for the Cultivation of Science,
Jadavpur, Calcutta 700 032, India.}

\date{\today}

\maketitle

\begin{abstract}
Based on a Fokker-Planck description of external Ornstein-Uhlenbeck noise
and cross-correlated noise processes driving a dynamical system we examine the 
interplay of the properties of noise processes and the dissipative characteristic
of the dynamical system in the steady state entropy production and flux. Our
analysis is illustrated with appropriate examples.
\end{abstract}

\pacs{PACS number(s) : 05.45.-a, 05.70.Ln, 05.20.-y}

\section{Introduction}
A dynamical system in contact with a reservoir has been a subject of wide 
attention in dissipative dynamics and irreversible thermodynamics.
The focal theme lies on the possible link between the rate of phase space volume 
contraction and the thermodynamically inspired quantities like entropy
production, entropy flux, Onsager coefficients etc. 
\cite{hol,leb,ruelle,jou,gaspard,bb3,cohen,pat,bb1,bb2,sc1}.
While on the other
hand it has been argued that the entropy production is related to the intrinsic
properties of phase space structure of the dynamical systems through the 
Lyapunov exponents \cite{bb3,cohen,pat,bb1,bb2,sc1}, the traditional wisdom asserts that entropy production
in a class of thermostatted Hamiltonian system
is defined \cite{ruelle} as the work per unit time (in the leading order) done on the system
by an external constraint under nonequilibrium steady state condition. Recently
based on a Markovian description of a stochastic process 
Daems and Nicolis \cite{nicolis} have
critically analysed the two aspects from the consideration
of an information entropy balance equation.

The object of the present paper is to extend the treatment to colour \cite{hr}
and cross-correlated noise processes \cite{jz,jz3} and to search for an appropriate
signature of an intrinsic interplay between the noise properties 
of these processes and the dissipative characteristics of the dynamical system
in the steady state entropy production and flux. We specifically consider
the overall system to be open, i. e. , the noises
are of external origin such that they do not, in general, satisfy 
fluctuation-dissipation (F-D) relations. Whenever possible we allow 
ourselves to make a fair comparison with the standard results for closed
systems.

The organisation of the paper is as follows: In Sec. II we 
consider two types of external, stationary and Gaussian noise processes
namely, the Ornstein-Uhlenbeck and cross-correlated noise processes in terms
of a Fokker-Planck description and set up an entropy balance equation to
identify the drift term which reveales that in addition to dissipation 
constant it contains the
essential properties of noise processes. Sec. III is devoted to explicit
examples to calculate the entropy production. The paper is concluded
in Sec. IV.

\section{The noise processes and entropy production}
\subsection{Fokker-Planck description}

\subsubsection{External Ornstein-Uhlenbeck noise processes}

The Langevin equations of motion in phase space for an N-degree-of-freedom 
system which is driven by the
external colour noise process $\eta_i$ can be written as

\begin{eqnarray}
\dot {q}_i & = & \frac{\partial H}{\partial p_i} = p_i \; \;, \nonumber \\
\dot {p}_i & = & 
-\frac{\partial H}{\partial q_i} -\gamma_i p_i +\eta_i, \; \; 
\; \; \; \; i = 1 \cdots N.
\label{e1}
\end{eqnarray}

\noindent
$N$ is the number of degrees of  freedom of the system. $\gamma_i$ is the
damping constant for i-th degree of freedom. 
$q_i, p_i$ are the corresponding co-ordinate and the
momentum, respectively. 
While the presence of $\gamma_i$ imparts a dissipative character in the
dynamics, the stochastic forcing $\eta_i$ ensures a canonical
distribution at equilibrium when the fluctuation-dissipation relation
is satisfied. $H$ is Hamiltonian 
of an initially the conservative system and is given by
\begin{equation}
H = \sum_{i=1}^N \frac{p_i^2}{2 } +V(\{q_i\}, t) \; \;.
\end{equation}

\noindent
The masses of all the degrees of freedom have been set to unity.
$V(\{q_i\}, t)$ is the potential of the Hamiltonian system.

The term $\eta_i$ in Eq. (1) refers to an external, Gaussian colour noise
for the i-th degree of freedom  and follows the two time correlation function

\begin{equation}
\langle \eta_i(t) \eta_i(t') \rangle = \frac{D_i^0}{\tau_i}  
e^{-\frac{|t-t'|}{\tau_i}}
\end{equation}

\noindent
where $\tau_i$ is the correlation time and $D_i^0$ is the noise strength.
The time evolution of $\eta_i$ can be conveniently expressed in terms of 
the white noise process $\zeta_i (t)$-for the i-th component  

\begin{eqnarray*}
\dot{\eta_i} & = &-\frac{\eta_i}{\tau_i} + \frac{\sqrt{D_i^0}}{\tau_i} \zeta_i 
\end{eqnarray*}
\begin{eqnarray}
\langle \zeta_i(t) \zeta_i(t') \rangle & = & 2 \delta (t-t') \nonumber\\
\langle \zeta_i\rangle & =& 0 \; \;.
\end{eqnarray}

In case there exists no fluctuation-dissipation relation between $\gamma_i$ and $\eta_i$
the system described by the Eq. (1) is sometimes termed as 
thermodynamically open \cite{lin}.

Eq.(4) implies that $\eta_i$ can be treated as a phase space variable on the
same footing as $q_i$, $p_i$. Thus the original $2N$ dimensional stochastic
system (1, 3) now becomes a $3N$ dimensional Markovian process where Eq. (1)
and  (4) are written in a compact form

\begin{equation}
\dot{X_i} = F_i(X) + \zeta_i
\end{equation}
where
\begin{eqnarray*}
X_i \left\{
\begin{array}{l}
= q_i \; \; \; {\rm for} \; i = 1, \cdots , N \\
= p_i \; \; \; {\rm for} \; i = N+1, \cdots , 2N \\
= \eta_i \; \; \; {\rm for} \; i = 2N+1, \cdots , 3N \\
\end{array}\right.
\end{eqnarray*}
\begin{eqnarray*}
F_i & = & X_{i+N} \; \; \; {\rm for} \; i = 1, \cdots , N  \nonumber\\
F_i & = & - \frac{\partial V(X_1, \cdots, X_N)}{\partial X_i} -\gamma_i X_i  
+ X_{i+N} \; \; \; {\rm for} \; i = N+1, \cdots , 2N   \nonumber\\
F_i & = & -\frac{X_i}{\tau_i} + \frac{\sqrt {D_i^0}}{\tau_i} \zeta_i 
\; \; \; {\rm for} \; i = 2N+1, \cdots , 3N
\end{eqnarray*}

\noindent
and

\begin{eqnarray}
\langle \zeta_i(t) \zeta_i(t') \rangle = 0 \; \; \; {\rm for} \; i = 1, \cdots , 2N \nonumber\\
\langle \zeta_i(t) \zeta_i(t') \rangle = 2 \delta (t-t') \; \; \; {\rm for} \; i = 2N+1, \cdots , 3N
\end{eqnarray}

The Fokker-Planck equation \cite{hr} corresponding to Langevin Eq.(5) can be written as

\begin{equation}
\frac{\partial P(X, t)}{\partial t} = - \sum_{i=1}^{3N} \frac{\partial}{\partial X_i}(F_i P)
+ \sum_{i=2N+1}^{3N} D_i
\frac{\partial^2 P}{\partial X_i^2 } \; \; \;,
\label{e7}
\end{equation}

\noindent
where $D_i = \frac{D_i^0}{\tau_i^2}$.

$P(X, t)$ is the extended phase space probability distribution function. The
extension is due to the inclusion of $N$ noise variables due to 
the external agency as phase variables. We conclude by pointing 
out that the above formulation contains the thermodynamically closed
system as a special case where the internal noise strength $D_i^0$ is related to 
dissipation $\gamma_i$
through the relation $D_i^0 = \gamma_i kT$, where $T$ refers to the equilibrium
temperature of the reservoir. 

\subsubsection{Cross-correlated noise processes}

Next we consider a dynamical system driven by both additive and 
multiplicative noise processes $\eta_i$ and $\zeta_i$,
respectively. The Langevin equation for this process, in general, can be written as

\begin{equation} 
\dot{X_i} = L_i(\{X_i\}, t) + g_i(X_i) \zeta_i +\eta_i \; \; \; i=1, \cdots ,N
\end{equation} 

\noindent
$L_i$ contains the dissipative term as well as the external applied 
deterministic force, if any. $g_i(X_i)$ is the 
coupling between the system and the multiplicative process. 
$\zeta_i$ and $\eta_i$ are white, Gaussian
noise processes with the following correlation between them;

\begin{eqnarray}
\langle \zeta_i(t) \zeta_j(t') \rangle 
& = & 2 D_{ij}' \delta (t-t') \delta_{ij} \nonumber\\
\langle \eta_i(t) \eta_j(t') \rangle & = &
2 \alpha_{ij} \delta (t-t') \delta_{ij} \nonumber\\
\langle \zeta_i(t) \eta_j(t') \rangle & = &
\langle \zeta_i(t') \eta_j(t) \rangle =
2 \lambda_{ij} \sqrt{D_{ij}'\alpha_{ij}} \delta (t-t') \delta_{ij}
\end{eqnarray}

\noindent
$D_{ij}'$ and $\alpha_{ij}$ correspond to the strength of multiplicative 
and additive noises, respectively. $\lambda$ represents the cross correlation 
between them with the limit $0\le \lambda \le 1$. The cross correlation between
these noise processes is known to cause symmetry breaking leading to 
non-equilibrium phase transitions \cite{jz} in spatially extended systems and generate
interesting ratchet motion \cite{jz3} in systems with symmetric potential under
isothermal condition.

The Fokker-Planck equation corresponding to Langevin Eq.(8) can be written as

\begin{equation}
\frac{\partial P(X)}{\partial t} = - \sum_{i=1}^{N} \frac{\partial}{\partial X_i}(F_i P)
+ \sum_{i=1}^{N} D_i
\frac{\partial^2 P}{\partial X_i^2 } \; \; \;.
\label{e10}
\end{equation}

\noindent
where the drift for the $i$-th component $F_i$ is

\begin{equation}
F_i = L_i(\{X_i\}, t) + \nu \left[ D_{ii}' \frac{\partial g_i(X_i)}{\partial X_i} 
g_i(X_i) + \lambda_{ii} \sqrt{\alpha_{ii} D_{ii}'}\right] + 
D_{ii}' \frac{\partial g_i^2(X_i)}{\partial X_i} +
2 \lambda_{ii} \sqrt{\alpha_{ii} D_{ii}'} \frac{\partial g_i(X_i)}{\partial X_i}
\label{e11}
\end{equation}

\noindent
$\nu =1$ stands for the Stratonovich and $\nu=0$ for the Ito convention, respectively.
Diffusion coefficient $D_i$ within small noise approximation can be written as

\begin{equation}
D_i = \alpha_{ii} + D_{ii}' g_i^2(X_{ie}) + 2 \lambda_{ii} \sqrt{\alpha_{ii} D_{ii}'} g_i(X_{ie})
\label{e12}
\end{equation}

\noindent
where $e$ in $X_{ie}$ refers to the steady state value of $X_i$, i. e., $X_{ie}$
is a solution of

\begin{equation}
F_i(\{X_i\}) = 0 \; \; \; i = 1, \cdots , N 
\label{e13}
\end{equation}

The choice of specific forms of nonlinearity in $L_i(\{X_i\}, t)$ results
in typical features of nonequilibrium phase transitions in model systems.
For the present purpose, however, we retain a general structure for
the rest of the treatment.

\subsection{Information Entropy production}

Information entropy $S$ is formally defined in terms of the phase space
distribution function $P(X, t)$ through the well-known relation 

\begin{equation}
S = -\int dX P(X,t) \ln P(X,t) \; \;.
\label{e14}
\end{equation}

The above definition allows us to have an evolution equation for entropy.
To this end we observe from Eqs.(\ref{e10}) (or (7)) and (\ref{e14}) that

\begin{equation}
\frac{dS}{dt} = -\int dX \left[ - \sum_i \frac{\partial}{\partial X_i}(F_i P)
+ \sum_i  D_i
\frac{\partial^2 P}{\partial X_i^2}
\right] \ln P \; \; \;.
\label{e15}
\end{equation}

Performing partial integration of the right hand side of the above Eq.(15)
and then dropping boundary terms (since the probability density tends
to zero as $|X|\rightarrow\infty$), one obtains the following form of
information entropy balance: 

\begin{equation}
\frac{dS}{dt} = \int dX P \nabla_X \cdot F
+ \sum_i D_i \int \frac{1}{P} 
\left(\frac{\partial P}{\partial X_i}\right)^2 \; \; \;.
\label{e16}
\end{equation}

The first term in (\ref{e16}) has no definite sign while 
the second term is positive 
definite because of positive definiteness of $D_i$. Therefore the 
second one can be identified as the entropy production ($\dot{S}_0$) \cite{nicolis}
\begin{equation}
\dot{S}_0 = \sum_i D_i \int \frac{1}{P_s} 
\left(\frac{\partial P_s}{\partial X_i}\right)^2 dX
\label{e17}
\end{equation}

\noindent
in the steady state. The subscript $s$ 
of $P_s$ refers to steady state. It is therefore evident from Eq.(\ref{e17}) that

\begin{eqnarray}
\dot{S}_{flux} & = & \int dX \; P_s(X) \; \nabla_X \cdot F \; \; 
 =  {\overline{\nabla_X \cdot F} } \nonumber\\
\dot{S}_0 & = & -\dot{S}_{flux} \; \;.
\label{e18}
\end{eqnarray}

\noindent
Note that since we consider the system to be dissipative 
$\overline{\nabla_X . F}$ is negative and therefore $\dot{S}_0$ turns out to
be positive.

\subsection{Influence of external perturbation}

It is now interesting to examine the entropy production when the dissipative
system is thrown away from the steady state due to an
additional weak applied force. To this end we consider the drift $F_1$ due to 
external force 
so that the total drift $F$ has now two contributions:

\begin{equation}
F(X) = F_0(X) + h F_1(X) \; \;.
\label{e19}
\end{equation}

When $h = 0$, $P = P_s$. The deviation of $P$ from $P_s$ 
in presence of nonzero
small $h$ can be explicitly taken into account once we make use of the identity
for the diffusion term \cite{nicolis}

\begin{equation}
\frac{\partial^2 P}{\partial X_i^2} = \frac{\partial}{\partial X_i}
\left[ P \frac{\partial \ln P_s}{\partial X_i}\right] +
\frac{\partial}{\partial X_i}\left[ P_s \frac{\partial}{\partial X_i}\frac{P}{P_s}\right] 
\; \; .
\label{e20}
\end{equation}

When $P = P_s$ the second term in 
(\ref{e20}) vanishes. In presence of additional
forcing the Eq.(\ref{e10}) becomes,
\begin{equation}
\frac{\partial P}{\partial t} = - \nabla_X . \psi P - h \nabla_X . F_1 P
+ \sum_i D_i \frac{\partial}{\partial X_i} \left( P_s \frac
{\partial}{\partial X_i}\frac{P}{P_s} \right)
\label{e21}
\end{equation}
where $\psi$ is defined as
\begin{equation}
\psi = F_{0} - \sum_i D_i \frac{\partial \ln P_s}{\partial X_i} 
\; \; .
\label{e22}
\end{equation}

Here we have assumed for simplicity that $D_i$ is not
affected by the additional forcing. The leading order influence
is taken into account through the additional drift term in Eq.(\ref{e21}).

Under steady state condition $(P = P_s)$ and $h = 0$, the second and 
the third 
terms in (21) vanish yielding

\begin{equation}
\nabla_X . \psi P_s = 0 \; \;.
\label{e23}
\end{equation}

It is immediately apparent that $\psi P_s$ refers to a current ${\cal J}$ 
where ${\cal J} = \psi P_s$. The steady state condition therefore reduces to
an equilibrium condition (${\cal J}=0$) if
\begin{equation}
\psi = 0 \; \; .
\label{e24}
\end{equation}
(In the next section we shall consider two explicit examples to show that $\psi=0$).
This suggests a formal relation between $F_0$ and $D_i$ as
\begin{equation}
F_0 = \sum_i D_i \frac{\partial \ln P_s}{\partial X_i}
\label{e25}
\end{equation}
where $P_s$ may now be referred to as the {\it equilibrium} density function
in phase space. 
$F_0$  contains dissipation constant $\gamma$. Depending on the problem it 
also depends on the correlation time $\tau_i$ of the colour noise or on the cross 
correlation $\lambda_{ii}$ between the noise processes.

To consider the information entropy 
balance equation in presence of external forcing
we first differentiate 
Eq.(\ref{e14}) with respect to time and use Eq.(\ref{e21}). 
Following Ref. \cite{nicolis} 
one can show that in the new steady state
(in presence of $h \ne 0$), the entropy production ($\dot{S}_h$) 
and the flux ($\dot{\Delta S}_{flux}$) like terms balance each other as follows:
\begin{equation}
\dot{S}_h = - \dot{\Delta S}_{flux}
\label{e26}
\end{equation}
with
\begin{equation}
\dot{S}_h = \sum_{i,j} {\cal D}_{ij} \int dX P\left(
\frac{\partial}{\partial X_i}
\ln \frac{P}{P_S}\right)^2 
\label{e27}
\end{equation}
and
\begin{equation}
\dot{\Delta S}_{flux} = h^2 \int dX \delta P \nabla_X . F_1 +h^2
\int dX \left( \sum_i F_{1i} \frac{\partial \ln P_S}{\partial X_i}\right)
\delta P \; \; \;.
\label{e30}
\end{equation}

\noindent
where we have put $h\delta P = P-P_s$.

In the following section we shall work out the specific cases to provide
explicit expressions for the entropy production 
and some related quantities due to external forcing for different kinds of open
systems mentioned in the last section.

\section{Applications}

\subsection{Entropy production in a system driven by an external colour noise}

To illustrate the theory we now consider a damped 
harmonic oscillator driven by an external, Gaussian Ornstein-Uhlenbeck
noise, $\eta_1$. The noise correlation of $\eta_1$ is given by Eq.(30).

\begin{eqnarray}
\dot{q_1} &=& p_1  \nonumber\\
\dot{p_1} &=& -\omega_0^2 q_1 -\gamma + \eta_1
\end{eqnarray}

\noindent

\begin{equation}
\langle \eta_1 (t) \eta_1 (t') \rangle = \frac{D^0}{\tau} 
e^{-\frac{|t-t'|}{\tau}}
\end{equation}

\noindent
where $\omega_0$ is the frequency of the oscillator.

To make notation consistent with Eq.(5) we would like to let $X_1$, $X_2$
and $X_3$ correspond to $q_1$, $p_1$ and $\eta_1$ respectively.

The relevant equations of motion are therefore as follows

\begin{eqnarray}
\dot{X_1} &=& F_1 = X_2  \; \; \; , \nonumber\\
\dot{X_2} &=& F_2 = - \omega_0^2 X_1 -\gamma X_2 + X_3  \; \; \; , \nonumber\\
\dot{X_3} &=& F_3 = - \frac{X_3}{\tau} + \frac{\sqrt{D^0}}{\tau} \zeta_3  \; \;,
\end{eqnarray}

\noindent
where $\zeta_3$ is a $\delta$-correlated noise

\begin{eqnarray*}
\langle \zeta_3 (t) \zeta_3 (t') \rangle = 2 \delta (t-t') \; \; \;.
\end{eqnarray*}

Therefore for the Langevin Eq. (31) the Fokker-Planck Eq.(7) becomes

\begin{eqnarray}
\frac{\partial P}{\partial t} & = & -X_2 \frac{\partial P}{\partial X_1}
+ \frac{\partial}{\partial X_2} (\omega_0^2 X_1 + \gamma X_2 -X_3) P +
+\frac{1}{\tau} \frac{\partial}{\partial X_3} (X_3 P) + 
\frac{D^0}{\tau^2} \frac{\partial^2 P}{\partial X_3^2} 
\end{eqnarray}

We now use the following transformation
\begin{equation}
U=a X_1+ b X_2 +X_3 \; \; \;,
\end{equation}
where $a$ and $b$ are constants to be determined.

Then under steady state condition Eq.(32) reduces to the following form:
\begin{equation}
\frac{\partial}{\partial U}(\Gamma U)P_s + D_s
\frac{\partial^2 P_s}{\partial U^2} = 0  \; \; ,
\end{equation}

where 
\begin{eqnarray}
D_s = \frac{D^0}{\tau},
\end{eqnarray}
and
\begin{eqnarray}
\Gamma U = -a X_2 + b \omega^2 X_1+ b \gamma X_2 -b X_3 + \frac{X_3}{\tau}   \; \; .
\end{eqnarray}

Here $\Gamma$ is again a constant to be determined. Putting (33) in
Eq. (36) and comparing the coefficients of $X_1$, $X_2$ and $X_3$ we find

\begin{eqnarray}
\Gamma a = -\omega_0^2 b \; \; \; \; , \nonumber
\; \; \; \; \Gamma b = -a + b \gamma  \; \;. 
\end{eqnarray}
\noindent
and

\begin{equation} 
\Gamma = -b + \frac{1}{\tau}
\end{equation}

The physically allowed solutions for $a$, $b$ and $\Gamma$ are as follows ;

\begin{eqnarray*}
a = \frac{1}{2}(-\frac{\gamma}{2}+\frac{1}{\tau} -\frac{1}{2}
\sqrt{\gamma^2 -4 \omega_0^2})(-\gamma-\sqrt{\gamma^2-4\omega_0^2})
\end{eqnarray*}

\begin{eqnarray*}
b = -\frac{\gamma}{2}+\frac{1}{\tau} -\frac{1}{2}
\sqrt{\gamma^2 -4 \omega_0^2}
\end{eqnarray*}

\noindent
and

\begin{equation}
\Gamma = -\frac{\gamma}{2}+\frac{1}{2}
\sqrt{\gamma^2 -4 \omega_0^2}
\end{equation}

The stationary solution of (34) $P_s$ is then given by
\begin{equation}
P_s = N_s e^{-\frac{\Gamma U^2}{2 D_s}} \; \;.
\end{equation}

Here $N_s$ is the normalization constant. By virtue of (39) $\psi$ 
corresponding to Eq.(22) is therefore

\begin{equation}
\psi = \Gamma U - D_s \frac{\partial \ln P_s}{\partial U} = 0 \; \;.
\end{equation}

Since $\psi P_s$ defines a current, $P_s$
defines a zero current situation or an equilibrium condition.
The equilibrium solution $P_s$ from (39) can now be used to calculate the
steady state entropy production as given by Eq.(17). We thus have

\begin{equation}
\dot{S}_0={\cal D}_s\int_{-\infty}^\infty  \frac{1}{P_s}
\left(\frac{\partial P_s}{\partial U}\right)^2 dU \; \; .
\end{equation}

Explicit evaluation shows
\begin{equation}
\dot{S}_0 = \Gamma \; \; ,
\end{equation}

\noindent
where $\Gamma$ is given by Eq.(38). Thus at equilibrium the entropy production
is inversely proportional to relaxation time of the process.

We now introduce an additional weak forcing in the dynamics. This may
achieved by adding a constant external force field $f_c$ in the dynamics.
Eq. (31) then  becomes

\begin{eqnarray}
\dot{X_1} &=& X_2  \nonumber\\
\dot{X_2} &=& - \omega_0^2 X_1 -\gamma X_2 + f_c + X_3  \nonumber\\
\dot{X_3} &=& - \frac{X_3}{\tau} + \frac{\sqrt{D^0}}{\tau} \zeta_3
\end{eqnarray}

Then the non-equilibrium situation (due to additional forcing, $F_{12}=f_c$)
corresponding to Eq.(43) is governed by

\begin{eqnarray}
\frac{\partial P}{\partial t} & = & -X_2 \frac{\partial P}{\partial X_1}
+ \frac{\partial}{\partial X_2} (\omega_0^2 X_1 + \gamma X_2 -X_3) P +
-\frac{\partial}{\partial X_2} (f_c P)+\frac{1}{\tau} \frac{\partial}{\partial X_3} (X_3 P) + 
\frac{D^0}{\tau^2} \frac{\partial^2 P}{\partial X_3^2} 
\end{eqnarray}

Using the transformation (33) again in Eq.(44) we have

\begin{equation}
\frac{\partial}{\partial U}(\Gamma U)P_s -\frac{\partial}{\partial U} F_u P+ D_s
\frac{\partial^2 P_s}{\partial U^2} = 0  \; \; ,
\end{equation}

\noindent
$\Gamma$ and other constants are given by the Eq. (38). Here $F_u$ is

\begin{equation}
F_u = b f_c
\end{equation}

The now stationary solution of (45) in presence of external forcing 
is now given by,

\begin{equation}
P'_s=N'e^{-\frac{\Gamma}{2D_s}[ U^2-\frac{2F_u U}{\Gamma}]} \; \; ,
\end{equation}
where $N'$ is the normalization constant.

We are now in a position to calculate the steady state entropy flux
($\dot{\Delta S}_{flux}$) due to external forcing ($h \ne 0$) from Eq.(28)

\begin{equation}
\dot{\Delta S}_{flux}=\int dX\delta P \nabla . F_1+ \int dX
\left(\sum_i F_{1i}\frac{d\ln P_s}{dX_i}\right) \delta P \; \;,
\end{equation}

\noindent
putting $h =1$.

The components of $F_1$ in $U$-space can be identified as

\begin{equation}
F_{11} = F_u \; \; \; {\rm and} \; \;  \nabla_U \cdot F_1 = 0
\end{equation}

$ \delta P=P'_s-P_s$ denotes the deviation from the initial equilibrium
state due to external forcing. For normalized probability functions $P'_s$
and $P_s$ the first integral in (48) vanishes.
Thus the entropy production at steady state due to weak forcing is given by

\begin{eqnarray*}
\dot{S}_h & = & -\dot{\Delta S}_{flux} \nonumber \\
& = & \frac{\Gamma}{D_s} \int F_u U \delta P dU  \nonumber\\ 
\end{eqnarray*}

Making use of the definition of $\delta P$ and integrating 
explicitly we obtain

\begin{equation}
\dot{S}_h =  \frac{b^2 f_c^2}{D_s}
\end{equation}

Putting $D_s$ from Eq.(35) and $b$ from (38) we obtain

\begin{equation}
\dot{S}_h = \frac{\left[4 -4 \gamma \tau + \gamma^2 \tau^2 
+ \tau^2(\gamma^2-4 \omega_0^2) -2 \tau \sqrt{\gamma^2 -4 \omega_0^2}( 
2 - \gamma \tau)\right] f_c^2}{4 D^0}
\end{equation}

We now examine specifically the following two limits:

\noindent
(i) In the Markovian limit $\tau \rightarrow 0 $  the above expression
reduces to the following form:

\begin{equation}
\dot{S}_h = \frac{f_c^2}{D^0}
\end{equation}

For the closed thermodynamic system $D^0 = \gamma kT$  which reduces the above
expression to the standard result for entropy production of irreversible 
thermodynamics for Brownian oscillator.

\noindent
(ii) Next we consider an interesting limiting case $\omega_0 \rightarrow 0$,
which implies that for a free Brownian particle we have

\begin{eqnarray*}
a = 0 \; \; \;, b = \frac{1-\gamma \tau}{\tau} \; \; \;  {\rm and} \; \; \; \Gamma = \gamma 
\end{eqnarray*}

\begin{equation}
\dot{S}_h = \frac{(1- \gamma \tau)^2 f_c^2}{D^0}
\end{equation}

The above expression depicts an interplay of the dissipation constant $\gamma$
of the system and the correlation time $\tau$ of the noise in determining the 
entropy production. Two different cases are noteworthy;

\noindent
(a) $\gamma \tau < 1$ or $\tau < \frac{1}{\gamma}$:

When relaxation time of the system greater than correlation time of external 
noise the entropy production $\dot{S}_h$ decreases with increase of
$\tau$ until $\tau \ge \frac{1}{\gamma}$.

\noindent
(b) $\gamma \tau > 1$ or $\tau > \frac{1}{\gamma}$:

The entropy production $\dot{S}_h$ increases with increase of $\tau$
until $\tau > \frac{1}{\gamma}$. It is interesting to note that
in the limit $\gamma \tau = 1$ entropy production is zero. A plot of entropy
production in the steady state vs correlation time therefore exhibits a minimum 
(See Fig. 1). It is thus apparent that in presence in presence of the nonequilibrium
constraint the properties of noise processes as well as the dynamic characteristic
of the system are important for entropy production.

\subsection{Entropy production in a cross-correlated noise driven system}

We now turn to the second case where a simple dissipative system is 
driven by both additive and multiplicative noises.

\begin{equation}
\dot{X_1} =  - \gamma X_1 -\zeta_1 X_1 +\eta_1
\end{equation}

Here $L_1$ in Eq.(8) corresponds to $-\gamma X_1$. The correlation between
the noise processes are given by,

\begin{eqnarray}
\langle \zeta_1(t) \zeta_1(t') \rangle &=& 2 D_{11}' \delta (t-t')  \nonumber\\
\langle \eta_1(t) \eta_1(t') \rangle &=& 2 \alpha_{11} \delta (t-t') \nonumber\\
\langle \zeta_1(t) \eta_1(t') \rangle &=& \langle \zeta_1(t') \eta_1(t) \rangle =
2 \lambda_{11} \sqrt{D_{11}'\alpha_{11}} \delta (t-t') \; \;, 0\le \lambda_{11} \le 1
\end{eqnarray}

\noindent
$\lambda_{11}$ denotes the cross-correlation between the two noise processes.

Eq.(10) for this system reduces to

\begin{equation}
\frac{\partial P(X_1)}{\partial t} = -\frac{\partial (F_1 P)}{\partial X_1}
+ D_1 \frac{\partial^2 P}{\partial X_1^2}
\end{equation}

\noindent
where the drift term is

\begin{equation}
F_1 = -(\gamma + 2 D_{11}' -\nu) X_1 + (2-\nu) \lambda_{11} \sqrt{D_{11}' \alpha_{11}}
\end{equation}

\noindent
and

\begin{equation}
D_1 = D_{11}' X_{1e}^2 -2 \lambda_{11} \sqrt{D_{11}' \alpha_{11}} X_{1e} + \alpha_{11}
\end{equation}

\noindent
where,

\begin{equation}
X_{1e} = \frac{(2-\nu) \lambda_{11} \sqrt{D_{11}' \alpha_{11}}}{\gamma + 2 D_{11}' -\nu}
\end{equation}

Making use of steady state value of $X_1$, i. e. $X_{1e}$ in Eq.(58) we obtain the
following constant diffusion coefficient in the weak noise limit

\begin{equation}
D_1 = \frac{\left[\alpha_{11} \gamma^2 + (2-\nu) D_{11}' \alpha_{11} \{(2-\nu)D_{11}'
+2\gamma -2 \gamma \lambda_{11}^2 -\lambda_{11}^2 (2-\nu) D_{11}' \} \right]}
{\Gamma'^2}
\end{equation}

\noindent
where

\begin{equation}
\Gamma' = \gamma + 2 D_{11}' -\nu
\end{equation}

Now the stationary solution of Eq.(56) is given by

\begin{equation}
P_s=N_1 e^{-\frac{\Gamma'}{2D_1}[ X_1^2-2\frac{2lX_1}{\Gamma'}]} \; \; ,
\end{equation}

\noindent
where $N_1$ is the normalization constant.

$l$ is given by

\begin{equation}
l=(2-\nu)\lambda_{11} \sqrt{D_{11}' \alpha_{11}}
\end{equation}

Putting (62) in Eq.(22) one may show as before that

\begin{equation}
\psi = 0
\end{equation}

Thus $P_s$ is an equilibrium probability distribution function.

Using Eq.(62) in Eq.(17) we obtain the standard expression for entropy production
at equilibrium

\begin{equation}
\dot{S}_0 =\Gamma'
\end{equation}

As before $\Gamma'$ is a negative divergence of the drift term in Eq.(61).
Eq.(65) carries same message as in Eq. (42) but for a different  
system. It is apparent that the cross correlated diffusion coefficient
$D_{11}'$ between the noise processes is as important as the dissipation factor
$\gamma$ that determines the steady state entropy production.

To study the effect of additional weak forcing on the stationary system we again
add a constant field of force $f_e$ in the Eq.(54). Due to the additional
forcing ($F_{11} = f_e)$ in Eq. (54) we have

\begin{equation}
\dot{X_1} =\gamma X_1 -\zeta_1 +\eta_1 +f_e
\end{equation}

Then the non-equilibrium situation corresponding to Eq.(66) is given by

\begin{equation}
\frac{\partial P}{\partial t} = \frac{\partial \Gamma' X_1}{\partial X_1}
-F_c' \frac{\partial P}{\partial X_1} +D_1 \frac{\partial^2 P}{\partial X_1^2}
\end{equation}

\noindent
where,

\begin{equation}
F_c' = f_e +l
\end{equation}

Using the stationary solution of Eq.(67) in Eq.(28) as in the previous section we
obtain the expression for entropy production in the steady state

\begin{equation}
\dot{S}_h = \frac{\left[\gamma^2 +(2-\nu)^2 D_{11}'^2 + 2\gamma (2-\nu) D_{11}'\right] f_e^2}
{\left[\alpha_{11} \gamma^2 +(2-\nu) D_{11}' \alpha_{11} \{(2-\nu)D_1 +2\gamma
-2\gamma \lambda_{11}^2 -\lambda_{11}^2(2-\nu) D_{11}'\}\right]}
\end{equation}

One may recover the standard results for a closed system by switching off the 
multiplicative noise ($D_{11}' = 0$) and implementing fluctuation-dissipation
relation $\alpha_{11} = \gamma kT$ in Eq.(69). We then obtain

\begin{equation}
\dot{S}_h = \frac{f_e^2}{\alpha_{11}} \; \; \; =\frac{f_e^2}{\gamma k T}
\end{equation}

Eq. (69) implies that for finite $D_{11}'$ entropy production is an increasing 
function of the cross correlation (i.e $\lambda_{11}$) between the two noise
processes.

\section{Conclusions}
In this paper we have examined the role of noise properties of stochastic 
processes in entropy production under a steady state condition. As specific
cases we have considered Ornstein-Uhlenbeck noise with finite correlation 
time and cross-correlated noises driving the dynamical system. Based on
an information entropy balance equation we have shown that the entropy production
and flux like terms not only depend on the dissipative characteristics of the dynamics
of the phase space of the dynamical system, particularly, the rate of phase space volume contraction,
but also on the correlation time and strength of cross correlation of the noises.
Since the steady state entropy production is identified as a drift term in the Fokker-Planck
description in the present formalism and the correlation time or the strength
of cross-correlated noises make their presence felt in this term, it is not
difficult to trace the origin of the role of interplay of dissipation and
the properties of the noise processes.
In view of the fact that the Ornstein-Uhlenbeck noise processes or the 
cross-correlated
noise processes are commonly occurring situations in condensed matter physics
and chemistry, we hope that the present analysis will be useful in
irreversible thermodynamics in relation to dynamical systems, in general.

\acknowledgements
S. K. Banik is indebted to the Council of Scientific and Industrial
Research for partial financial support.

\begin{figure}
\caption{
Plot of entropy production $\dot{S}_h$ vs correlation time $\tau$ 
for the Eq.(53) using $\gamma =1.0$, $f_c = 1.0$ and $D^0 = 1$ (Units are arbitrary).
}
\end{figure}

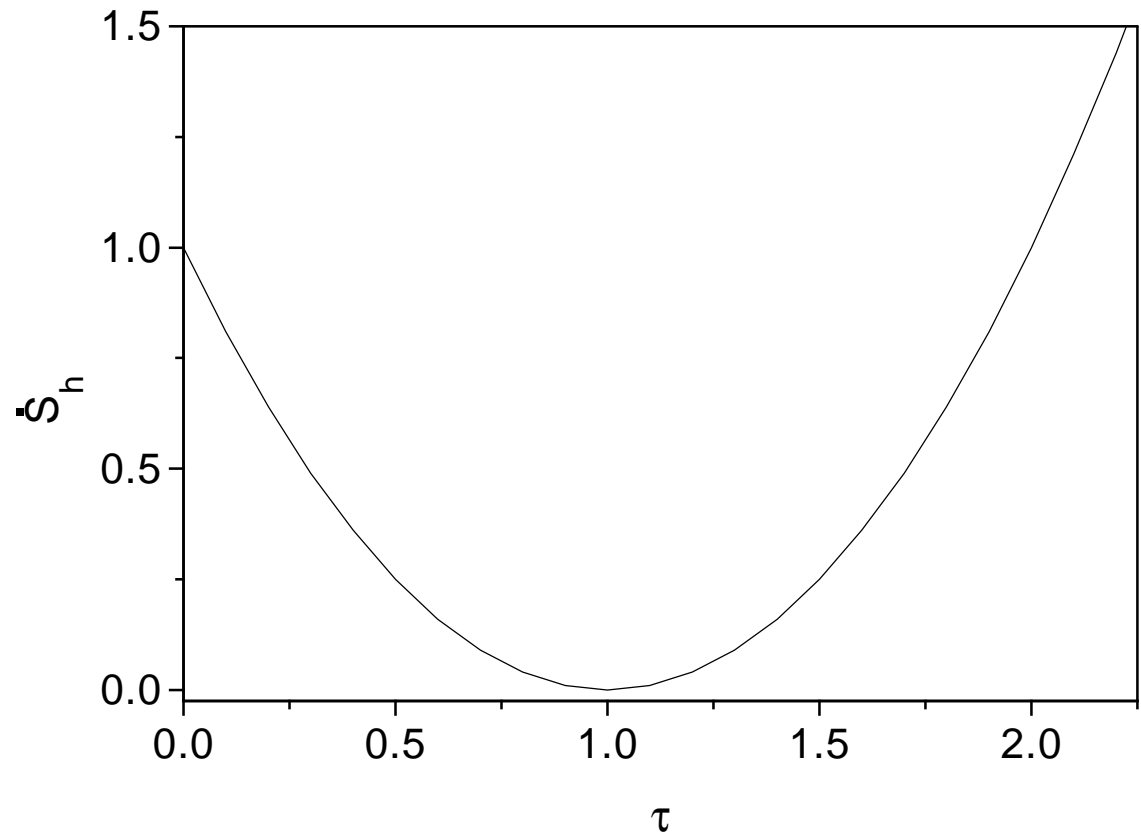
\end{document}